# on spherically-symmetric accretion by

# a collisionless polytrope

B. M. Lewis

Arecibo Observatory

blewis@naic.edu

Arecibo Observatory
HC3  Box 53995
Arecibo, PR00612

**Abstract**

An isolated, spherically-symmetric, self-gravitating, collisionless system is always a polytrope when it reaches equilibrium (Nakamura 2000). This strongly suggests as a corollary, however, that the *same* polytrope dominates its precursor states, since the dynamical equations for its constituents can be time-reversed. Moreover this assumption, which precludes a polytrope from ever accreting 100% of the mass from an infalling shell, as a subsequent state will eventually be a polytrope, is confirmed now by our finding that a collisionless polytrope cannot accrete 100% of an infalling shell while simultaneously guaranteeing that the entropy of the Universe as a whole increases. These strictures are only evaded by the shedding of some mass. A polytrope must lose mass to gain mass.

We deduce from the time-reversible property of a collisionless polytrope that the scalar sum, $P$, over constituent momenta in its rest frame is an independent state variable that is conserved with respect to its surface radius through interactions between a polytrope and an infalling shell. This new constraint, together with conservation of energy, enables us (i) to show that an isolated polytrope is indeed stable against spherically-symmetric mass-loss, which is the essential content of our initial assumption; (ii) to calculate both the velocity and the fraction of infall-mass returned to infinity, provided the "accretion law" between the change in mass and surface radius is specified. Numerical results confirm a frequent empirical finding (Livio 2000) that the velocity of a mass outflow is of the same order of magnitude as the escape velocity from the system.

gravitation

stars: mass-loss           stars: AGB and post-AGB

galaxies: evolution  --   galaxies: kinematics and dynamics

**Introduction:**

The program initiated by Lynden-Bell (1967), to understand the ordered distribution of stars in globular clusters and elliptical galaxies through an application of statistical mechanics to collisionless systems, was completed by Nakamura (2000). Lynden-Bell based his theory on conservation of phase-space volume, which he divides into small "microcells", and arranges into "macrocells". He then counts the number of combinations of macrocells that conserve the total energy of the system, expecting the equilibrium distribution to be the one that maximizes this number under the conservation laws. But this approach leaves a residual "velocity dispersion" problem, as the theory predicts that the equilibrium distribution has a superposition of Gaussian components with different velocity dispersions in the appropriate nondegenerate limit. By contrast Nakamura uses a Maximum Entropy approach, following Jaynes (1957), which maximizes the number of combinations of occupied microcells that are consistent with the constraints of conservation of phase-space and energy. Nakamura finds that the constituent particles of a collisionless, self-gravitating body have a single Gaussian velocity dispersion, but do not have equipartition of energy. We thus expect the mass-distribution of a self-gravitating, non-rotating, spherically-symmetric ensemble of collisionless particles to be an isothermal sphere, no matter whether such "particles" are stars, dark matter, or a mixture of both.

The dynamical equations for the constituents of a collisionless system can be time-reversed. This leads inexorably to an expectation that the *same* polytrope occurs among its past states, as any configuration assumed by such a system is either a polytrope, or in due course becomes a polytrope. Which in turn raises a question as to how such systems originate? Let us suppose for a moment that a polytrope is able to accrete 100% of the mass from an infalling spherical shell. Then, from Nakamura's result, its eventual state is again a polytrope. Whereupon the time-reversibility of the equations for its constituent parts would demonstrate that a polytrope is inherently unstable to spherically-symmetric mass-loss, with the likely rider that it progressively becomes ever more concentrated as it evaporates over time. The alternative assumption is to assert that an isolated polytrope is inherently stable against spherically-symmetric mass-loss, to which the ubiquity of globular clusters and elliptical galaxies may attest. In which case a polytrope could only increase in mass if some of the incoming mass is returned to infinity: time reversing the equations merely causes the diaspora of expelled mass to return to the system, to interact with it, before leaving again together with the whole of the originally accreted shell at a lower velocity. An isolated polytrope could indeed be stable.

What actually happens when a polytrope accretes a shell from infinity? Since none of the incoming energy can be radiated by a collisionless system, its binding energy must decrease when 100% of it is accreted, so the polytrope must expand. Section 2 shows that the entropy of the Universe as a whole will in general decrease as a consequence, though this outcome is easily bypassed if a fraction of the infalling mass returns to infinity. This is one of the reasons why accreting systems so often exhibit signs of ongoing mass-loss, which occurs in such diverse contexts as star-formation, long-period variables, cataclysmic variables, micro-quasars, and quasars. Section 3 explores a particular case for the accretion of mass by a polytrope wherein the net entropy per particle remains constant, to show that the entropy of the Universe increases. An additional relation is needed (besides conservation of energy) to determine both the fraction of the incoming mass returned to infinity and its velocity. Section 4, while building on Lewis (1997), argues that this relation is provided by conservation of the scalar sum over the constituent momenta of the system. Section 5 completes the exercise and generalizes to the full range of physically possible accretion relations.

## 2. The entropy of a polytrope

This section is concerned with the entropy change to the Universe as a whole induced by accretion of a spherical shell in free-fall from infinity onto a polytrope. Let the six-dimensional phase-space volume, $V_6$, of a polytrope of mass, $M$, composed of $N$ particles of average mass, $m$, so $N = M/m$, be divided into a set of microcells of equal volume. When each microcell has the same probability, $p_i$, of occupancy by a particle, the "information entropy" of the system is defined as the sum over all microcells

$$S_{poly} = -\sum_i p_i \ln(p_i) = -N \ln(p_i) \qquad (1),$$

following Jaynes (1980).

We need the phase-space volume, $V_6$, of a polytrope to progress. A polytrope of index n and surface radius $R_s$ has a gravitational potential energy (Chandrasekhar 1939, p101)

$$\Omega = -\left(\frac{3}{5-n}\right)\frac{GM^2}{R_S}.$$

As this is a self-gravitating body in equilibrium, it satisfies the Virial Theorem, so its total kinetic energy, $T$, is related to $\Omega$ via $2T + \Omega = 0$ : $T$ is therefore related to the escape velocity from the surface, $V_{esc}$, through

$$T = \tfrac{1}{2} M V_{rms}^2 = \tfrac{1}{2} M \lambda_n^2 V_{esc}^2 \quad ; \quad \lambda_n^2 = \frac{1}{2}\left(\frac{3}{5-n}\right) \qquad (2),$$

and the maximum velocity at its center through a constant, χ, such that

$$V_{max} = \chi \lambda_n V_{esc} \qquad (3),$$

so the six-dimensional phase-space volume is

$$V_6 = \left(\tfrac{4}{3}\pi V_{max}^3\right)\left(\tfrac{4}{3}\pi R_S^3\right) = \tfrac{16}{9}\pi^2 \chi^3 \lambda_n^3 \left(V_{esc} R_S\right)^3 = \tfrac{16}{9}\pi^2 \chi^3 \lambda_n^3 \left(2 G M R_S\right)^{3/2} \qquad (4).$$

When for convenience $V_6$ is divided into $2N$ microcells of volume

$$\Delta V_6 = V_6 / 2N = V_6\, m / 2M \qquad (5),$$

the probability of the i-th microcell containing a particle is just the ratio between the number of particles and the number of microcells, $p_i = N \Delta V_6 / V_6$ ($\equiv 0.5$). The resulting "entropy" follows from substituting $p_i$ into equation 1. Nevertheless the number of microcells into which we have chosen to divide $V_6$ is both arbitrary and sets the magnitude of our "entropy", so $S_{poly}$ is not in itself a physical quantity, nor is it identical to the thermodynamic entropy of the system, though its properties cause it to closely track it.

This entropy calculation is robust. Moreover equation 4 shows that $V_6$ decreases when a constant-mass polytrope is allowed to contract by radiating energy, so $p_i$ must increase when $\Delta V_6$ is constant. Thus the net entropy of a polytrope decreases as it contracts, and so must be compensated by the entropy gained by the rest of the Universe from its emitted photon shell. And vice versa: when a polytrope increases in size and/or mass, $V_6$ increases, as therefore does its entropy.

We seek next to quantify the entropy gain induced in a *collisionless polytrope* by the accretion of $dM$ in free-fall from infinity under conservation of mass and energy. Since Nakamura finds that the eventual equilibrium state of the merged entity is a polytrope, the change in its potential energy on addition of $dM$

$$\Delta \Omega = \left(\frac{3}{5-n}\right)\frac{G M^2}{R_S} - \left(\frac{3}{5-n}\right)\frac{G (M + dM)^2}{R_S} = -2\Omega\left(\frac{dM}{M}\right) \quad (6)$$

initially augments $T$, so
$$dT = 4T\left(\frac{dM}{M}\right) = 2\lambda_n^2 V_{esc}^2 \, dM \quad (7).$$

The original binding energy of $E = \Omega/2 = -T = -\frac{1}{2} M \lambda_n^2 V_{esc}^2$, becomes

$$E' = 2\lambda_n^2 V_{esc}^2 \, dM - \frac{1}{2} M \lambda_n^2 V_{esc}^2 = -\frac{1}{2} M \lambda_n^2 V_{esc}^2 (1 - 4\, dM/M) = -T',$$

where primed quantities denote parameters of the polytrope after it reaches equilibrium, while conserving energy and mass. But by definition

$$T' = \frac{1}{2}(M + dM)\lambda_n^2 V'^{2}_{esc} = \frac{1}{2} M \lambda_n^2 V_{esc}^2 (1 - 4\, dM/M).$$

On substituting for $V_{esc}$ and simplifying this becomes

$$R'_S (1 - 4\, dM/M) = R_S (1 + dM/M)^2 = R_S (1 + 2\, dM/M),$$

which on retaining first order terms simplifies to

$$dR_S = R'_S - R_S = 6 R_S (dM/M) \quad (8).$$

The change in phase-space volume, from differentiating equation 4, is

$$\left(\frac{dV_6}{V_6}\right) = \frac{3}{2}\left(\frac{dM}{M} + \frac{dR_S}{R_S}\right) \quad (9),$$

which on substituting from equation 8 gives

$$dV_6 = \tfrac{21}{2} V_6 \,(dM/M) \qquad (10).$$

This translates, via equation 5, to

$$dV_6 = \tfrac{21}{2} \Delta V_6 \,(2\,dN) \qquad (11)$$

and so adds 10.5 times as many microcells of phase-space per added particle as the original polytrope had. The resulting entropy change, after differentiating equation 1 and substituting from 5 & 11, is

$$dS_{poly} = dN \left( \tfrac{19}{2} - \ln p_i \right) \qquad (12).$$

There is thus an entirely predictable increase in the entropy of a polytrope when $dM$ completely merges with $M$, while conserving the system's total energy.

However the entropy content of the infalling shell, $dS_{in}$, is effectively lost in this merging process, as every microcell of infalling phase space is then for ever after empty, whereas $dS_{in}$ has a completely unconstrained magnitude relative to $dS_{poly}$. Thus infalling matter could either arrive in seconds or gradually over a Hubble time, while $dS_{poly}$ is the same in both cases. This *merging process* consequently lacks any constraints to guarantee

$$dS_{poly} \geq dS_{in} \,,$$

and so fails to satisfy our basic expectation that no physical process can cause the entropy of the Universe as a whole to decrease. We therefore conclude that at least one of its premises needs to be adjusted.

There are three solutions: (i) infalling mass merges with the polytrope, but leaves it perpetually out of equilibrium; (ii) all of the infalling mass is returned to infinity; or (iii) a portion of the infalling mass is returned to infinity so that the entropy of the Universe as a whole increases. The first possibility appears to be ruled out by Namakura's result as well as by the prevalence of polytropic mass-distributions in globular clusters and elliptical galaxies, while the second would make it difficult to accumulate mass into astronomical bodies without an accompanying dissipative process. This is particularly apposite in the case of rich clusters of galaxies, which are dominated gravitationally by dark matter while having an isothermal mass distribution. The third possibility is explored next.

## 3. isentropic accretion

The central problem identified in §2 is the impossibility of a fixed general relationship between the predictable entropy gain of an accreting polytrope, and that lost or subsumed from a completely merged infalling shell. The only viable alternative is to assume that accretion occurs in such a fashion that the entropy gain of the polytrope is smaller, a particularly simple *pro tem* choice being to adopt the "isentropic accretion law", which maintains the net entropy per particle in the polytrope constant, so

$$\left(\frac{dV_6}{V_6}\right) = \left(\frac{dM}{M}\right) \qquad (13).$$

Equations 9 & 13 then show

$$dR_s = -\tfrac{1}{3} R_s \left(dM/M\right) \qquad (14).$$

As $dM$ is positive for accretion, the sign in equation 14 signifies that the polytrope contracts under the isentropic constraint, in marked contrast to that of equation 8, while

$$dS_{poly} = -(dM/m)\ln[p_i] = -dN \ln[p_i] \quad (15)$$

is still positive. Yet any contraction of a stable polytrope releases an excess-to-virial-stability energy, $\Delta E$, that cannot be absorbed by the polytrope. Since a collisionless system has no dissipation mechanisms, $\Delta E$ has to be carried away in a particle flux, so a portion $\alpha\, dM$ of the infalling mass must return to infinity. This potentially solves the entropy problem of the infalling shell too, as the lower probability for occupancy $p_{i,out}$ of microcells in the decimated shell as it expands with $v(R_s) > V_{esc}$ increases its entropy.

The entropy of the infalling shell is $dS_{in} = -(1+\alpha)(dM/m)\ln[p_{i,in}]$, where $p_{i,in}$ is

$$p_{i,in} = (1+\alpha)\left(\frac{dM}{m}\right)\frac{\Delta V_{6,sh}}{16\pi^2 R_s^2 V_{esc}^3\, dt\, dv} = (1+\alpha)\left(\frac{dM}{m}\right)\left(\frac{p}{V_{esc}^3}\right) \quad (16).$$

When $\alpha\, dM$ of the infall-mass is expelled with a velocity $V_\infty > 0$, the probability of occupancy of a microcell in the outgoing shell of velocity $v(R_s)$ is

$$p_{i,out} = \left(\frac{\alpha\, dM}{m}\right)\left(\frac{p}{v^3}\right) \quad (17),$$

which on substituting equation 16 gives

$$p_{i,out} = \left(\frac{\alpha}{1+\alpha}\right)\left(\frac{V_{esc}}{v}\right)^3 p_{i,in} \quad (18).$$

The entropy increase of the outgoing shell is

$$\Delta S = (dM/m)\left\{\ln[p_{i,in}] - \alpha \ln\left[\left(\frac{\alpha}{1+\alpha}\right)\left(\frac{V_{esc}}{v}\right)^3\right]\right\} \quad (19),$$

which is demonstrably positive when the entropy scale for the accreted mass is set so that $p_{i,\,in} \equiv 1$. Thus equations 15 & 19 together show that the entropy of the Universe as a whole increases under the isentropic accretion of mass by a polytrope.

This result comes at the cost of introducing two parameters $\{\alpha \ \& \ v(R_s)\}$ for any specific choice of accretion "law", so two constraints are needed to evaluate them, of which one comes from conserving energy. It is worth remembering at this point that the complete accretion of the incoming shell is permitted by Nakamura's assumption that the only pertinent constraints are conservation of phase space and energy. Our analysis in §2, by negating this assumption, points to the existence of at least one additional state variable, which should provide a second constraint. This is related in most dynamical problems to momentum. The scalar sum $P$ over constituent momenta was proposed by Lewis (1997) as a state variable: it leads to $v(R_s) \approx 1.5\, V_{esc}$ and $\alpha \approx 0.82$ in the isentropic case. Section 4 outlines the background needed to deploy a scalar momentum constraint.

## 4. the scalar momentum constraint

Let $P$ be the scalar sum over the momenta of a polytrope's particles in its rest frame

$$P = M\, \bar{v} \qquad (20),$$

where the mean velocity, $\bar{v}$, is from equation 2

$$\bar{v} = \lambda V_{rms} = \lambda \lambda_n V_{esc}$$

and $\lambda = 0.921$ for a Gaussian velocity distribution. Thereupon the relation between $P$ and the total kinetic energy $T$ of a polytrope is

$$\frac{P^2}{2M} = \lambda^2 T \qquad (21),$$

which shows that $P$ is a statistical constant for a stable polytrope of constant $M$ and $R_s$ like $T$. $P$ & $T$ are closely inter-related, as any parameter concerned with the absolute magnitude of a particle's momentum, $\vec{p}_i$, is also related to $\vec{p}_i \bullet \vec{p}_i$, and thus to its kinetic energy. So the $P$ of a polytrope cannot in general change without a simultaneous change in $T$. Moreover, as $T$ is part of the "internal energy" of the system (e.g. Landau & Lifshitz 1960, p17), $P$ can likewise be considered its "internal momentum", of which more below. By differentiating equation 21, while holding the mass constant, we get

$$dP = \frac{P}{2T} dT \qquad (22):$$

hence the homologous contraction of a polytrope changes its $P$ in exact proportion to that in $T$. More generally the differentiation of equation 20 with respect to $M$ and $R_s$ gives

$$dP = \frac{3}{2} \bar{v} \, dM - \frac{\bar{v}}{2} \frac{M}{R_s} dR_s \qquad (23),$$

where the first term shows that the change in $P$ of the resulting polytrope is positive under the addition of mass, while the second shows it is positive for contraction (negative $dR_s$).

What happens to $P$ when a *collisionless* polytrope contracts homologously through $dR_s$? Clearly the momentum of every particle increases as it is accelerated by gravity towards the center, so the whole of the resulting change in the gravitational potential energy, $d\Omega$, is translated initially into a kinetic energy increase

$$dT = -2(T/R_s)dR_s = -(|\Omega|/R_s)dR_s = |d\Omega| \quad ,$$

with $P \to P + dP$ following equation 22. Since the Virial Theorem requires the $T$ of the polytrope *in equilibrium* at its new $R_s$ to increase by just $dT = |d\Omega|/2$, a *collisionless system* must first export $dT = |d\Omega|/2$ of mechanical energy to reach this state. Moreover, as every part of $d\Omega$ is linked to a proportionate quantity of the contraction-generated scalar momentum, the export of $dT$ is necessarily linked by equation 22 to the export of scalar momentum. On the other hand, in the absence of a mechanism for exporting energy (or scalar momentum), the polytrope returns to its prior size, a state that can also be reached by time-reversal. Hence a perturbed polytrope must have both the requisite mechanical energy and scalar momentum to return to equilibrium.

We thereby reach three important conclusions. Firstly, since the export of $dT$ of mechanical energy implicitly implies its passage through the surface radius, $R_s$, this understanding must also be applied to the linked export of $dP$. Thus interactions with polytropes can always be calculated via reference to the situation obtaining at $R_s$. Secondly the magnitudes of $dT$ & $dP$ exported by a polytrope are conserved relative to $R_s$, since the prior state is accessible by time-reversal. An alternative way of viewing this point is that no work is done on the system in exporting $dT$ & $dP$, so these quantities are conserved. Thirdly, since, by Nakamura's result, the equilibrium state of a self-gravitating system is a polytrope for which the equilibrium ratio between $T$ & $P$ is a given, a system must export whatever differential fraction of either is needed to reach equilibrium.

Four dicta govern collisionless self-gravitating systems. These are (1) their equilibrium state is a polytrope; (2) their dynamical equations are time-reversible; (3) under a homologous contraction they swap gravitational potential energy into kinetic energy in such a way that the ratio between $P$ & $T$ follows a definite relation; (4) given the absence of a mechanism for exporting energy or momentum, the future state of an *isolated* system is again that of a polytrope with the same state variables as the initial system. Thus any system which passes through a state that can be *completely* identified with one occurring during the homologous contraction of a polytrope, eventually returns to being a polytrope in equilibrium, and, moreover, had the *same* polytrope as a precursor.

Besides the excess-to-the-needs-of-virial-stability $dP$ generated by homologous contraction of a polytrope given by the second term of equation 23, we need the similar excess following on the complete merging of a mass increment, $dM$, into a polytrope at constant $R_s$. Its maximum size is limited in general to the scalar momentum generated by the free-fall of $dM$ in a spherical shell from infinity. This, from Lewis (1997), is

$$\Delta P = 2P \left( dM / M \right) = 2 \bar{v} \, dM \quad (24),$$

which is seen to be qualitatively appropriate as the addition of $dM$ to a polytrope requires $dP = \bar{v} \, dM$ to match the rest of it, and this can only be linked to half of the net change in gravitational potential energy. As the first term of equation 23 sets the change in the $P$ of a polytrope in equilibrium from this process at $1.5 \bar{v} \, dM$, the rest of the polytrope must need $0.5 \bar{v} \, dM$ to compensate for its increased mass. The excess-to-the-needs-of-virial-stability $dP$ generated by changes to $M$ and $R_s$ is therefore

$$dP = \frac{\bar{v}}{2} dM - \frac{\bar{v}}{2} \frac{M}{R_s} dR_s \quad (25),$$

while the excess-to-the-needs-of-virial-stability energy is

$$|d\Omega| / 2 = T \left[ 2 \left( \frac{dM}{M} \right) - \left( \frac{dR_s}{R_s} \right) \right] = dT \quad (26),$$

with the first term in each case coming from the complete merging of a shell at constant $R_s$, and the second from contraction at constant $M$.

When a polytrope contracts homologously by $dR_s$, the acceleration of every particle is strictly radially inward, as they move in a centrally symmetric potential: $dP$ therefore has a radially-directed character. It thus has the same nature as the radially-directed momentum flux $L/c$ commonly accorded a body's luminosity (Salpeter 1974; Knapp *et al.* 1982). Moreover when the momentum vector of every constituent particle in the rest frame of a polytrope, $\vec{p}_i$, is viewed as emanating from a common center, its "scalar momentum" $P$ becomes a summation over $4\pi$ steradians of the individual vectors, so though omni-directional in Cartesian space, it is radially-directed when viewed from a centrally-symmetric space. This is in accord with the evident algebraic additivity of $P$ as a polytrope contracts or expands homologously and $P \to P \pm dP$, which is also expected from the addition/subtraction of vectors having the same direction,

$$\vec{P} \to \vec{P} \pm d\vec{P}.$$

Nor is this surprising, as radial acceleration is the only kind that can occur in a centrally symmetric potential. Thus the radially-directed quantity $\vec{P}$, following equation 21, is effectively the "internal momentum" of the system.

Is the scalar momentum parameter $P$ anything more than a different parametrization of the kinetic energy? This is readily settled now by considering the change in $P$ putatively following on the complete merging of an infalling shell, when the accreting polytrope absorbs its energy. The consequent $dR_s$ from equation 8, on substitution into 25, gives

$$dP = -\frac{5}{2} \bar{v}\, dM \quad (27),$$

where the negative sign tells us that a deficit in $P$ is produced by the process. This shows that $P$ for the entire system is something more than a reformulated version of $T$. Though the equilibrium of a polytrope mandates a specific relationship between its $P$ & $T$ that is conserved by homologous operations, this is not conserved by polytrope-independent operations, such as the infall of a shell. Thus the deficit in $P$ follows from the total change in the gravitational potential energy (and scalar momentum) of the infalling shell being initially resident in just the shell's particles, whereas the polytrope must disperse this energy (and momentum) among its more slowly moving particles, before it can undergo an homologous expansion to reach the "energy-mandated" equilibrium of equation 8.

Let us suppose for a moment that the "merged system" can reach the equilibrium state defined by equation 8, and label it **B**. Then time-reversal must lead back to the prior state **A** at which the infalling shell has been completely merged. State **A** has the same total mass as **B**, with the mass distribution of a polytrope even while it has the same kinetic energy as would result from a homologous contraction from **B**. Nevertheless **A** differs from a state accessible by homologous contraction from **B** in having a different $P$. Further, the continuation of the time-reversed trajectory of **A** must lead it to eject a shell to infinity. Neither state **A**, nor the subsequent ejection of a shell, is a possible outcome on any dynamical trajectory for a polytrope undergoing homologous contraction from **B**. Moreover the sole pointer to the divergent fate of **A** is the "anomalous" magnitude of its $P$. We therefore conclude that $P$ is an independent state variable.

This brings us to our central thesis. We are led by the time-reversibility of the equations, and by our expectation that the only permissible processes are those that ensure that the consequent entropy change of the Universe as a whole has $dS \geq 0$, to believe that a polytrope cannot "manufacture" the requisite radially-directed momentum to compensate for the deficit of equation 27, by redistributing energy among its parts. By conserving the $P$ of the total system at the reference radius $R_s$, nature distinguishes between otherwise equivalent fates and so eliminates any possibility of "dynamical degeneracy", as would occur for instance if both the time-reversed solution from **B** to **A** and homologous contraction from **B** to **A** were simultaneously possible. In the present case this ensures that **B** is not accessible from **A**, so time-reversal would send a shell off to infinity again with $V_\infty = 0$. Moreover, with both initial and final states being polytropes, every possible bisymmetric accretion process must expel some fraction of the incoming mass to balance the ratio mandated by a polytrope between its $P$ & $T$. A *collisionless system therefore conserves its scalar (aka internal) momentum,* except for the entirely predictable changes due to motion in its centrally-symmetric potential. Whereupon equation 27 on its own suffices to rule out any possibility that a polytrope can completely absorb a shell.

Is a polytrope stable under the constraints of conserving energy and momentum? Section 1 shows that it is unstable when the only constraint is conservation of energy, so the present question is whether the addition of the principle of conserving scalar momentum guarantees its stability. This is demonstrated next.

The process of losing a spherically symmetric shell of mass $dM$ is equivalent to the exact inverse process to its accretion. Consequently whilst accretion generates surplus energy (equation 26) and scalar momentum (equation 25), these exact quantities, by time-reversal, must be supplied to effect the demerging of a shell. Let us consider a collisionless polytrope, which by definition is in equilibrium. Equation 24 sets the $dP$ generated by the infall of a shell at $2\bar{v}\,dM$, of which $1.5\bar{v}\,dM$ is retained by the polytrope and $0.5\bar{v}\,dM$ is exported as the polytrope returns to equilibrium. To lose $dM$ (so its velocity at infinity is zero) thus requires $dP = 0.5\bar{v}\,dM$ to be provided to the "demerging" shell. If for generality the outwardly directed velocity of $dM$ is $v(R_s)$,

$$dP_{demerge} = 0.5\,\bar{v}\,|dM| + |dM|\left(v(R_s) - V_{esc}\right) \quad,$$

which devolves to the expected limit when $v(R_s) \equiv V_{esc}$. Since $dP$ is supplied in this case by contraction of the residual polytrope, on substituting the second term of equation 25 we get the "momentum" equation for this process

$$0.5 \, \bar{v} \, |dM| + |dM| \left( v(R_s) - V_{esc} \right) = -0.5 \, \bar{v} \left( M/R_s \right) dR_s \; .$$

This can be rewritten, after substituting for $\bar{v}$ and setting $\xi = v(R_s)/V_{esc}$, as

$$|dM| \left\{ \lambda_n^2 + 2\left(\frac{\lambda_n}{\lambda}\right)(\xi - 1) \right\} = -\lambda_n^2 \left(\frac{M}{R_s}\right) dR_s \quad (28).$$

Likewise, to demerge a shell so its $V_\infty = 0$ requires the provision of energy equal to the shell's kinetic energy, which from equation 26 is $2T(|dM|/M)$. As this is also obtained from the contraction of the residual polytrope, the general energy equation, after using the second term of equation 26, is

$$2T\left(\frac{|dM|}{M}\right) + \frac{1}{2} |dM| \left( v^2(R_s) - V_{esc}^2 \right) = -\frac{1}{2} \lambda_n^2 V_{esc}^2 \left(\frac{M}{R_s}\right) dR_s \; ,$$

which on substituting $\xi$ and rearranging becomes

$$|dM| \left\{ 2\lambda_n^2 + (\xi^2 - 1) \right\} = -\lambda_n^2 \left(\frac{M}{R_s}\right) dR_s \quad (29).$$

Since the same contraction through $dR_s$ supplies both the requisite momentum and energy, the left hand sides of equations 28 & 29 on rearranging give the quadratic in $\xi$

$$\xi^2 - 2\left(\frac{\lambda_n}{\lambda}\right)\xi + 2\left(\frac{\lambda_n}{\lambda}\right) + \lambda_n^2 - 1 = 0 \quad (30),$$

where physically meaningful values for $\xi$ require $\xi \geq 1$. The roots of equation 30 for the isothermal polytrope (n = 3/2) with $\lambda$ = 0.921 are 0.7108 ± j 0.5873 show that it is **not possible** for an isothermal sphere to lose a spherically-symmetric (or indeed a bi-symmetric) shell of mass, while conserving both energy and scalar momentum. Thus a polytrope is inherently stable under these constraints.

## 5. accretion by a collisionless polytrope

The $R_s$ of a collisionless polytrope must in general change under the spherically symmetric accretion of mass, where the initially allowed range for $dR_s$ as a function of $dM$ stretches from that under conservation of energy (equation 8), down to at least the isentropic case of equation 14. While these solutions balance the energy equation, by using part of the incoming mass to carry away the excess contraction energy, this process inherently requires a positive (i.e. excess) $dP$ from equation 25 to operate. The range of possible solutions, after setting $dP = 0$ in equation 25, is therefore

$$-\tfrac{1}{3} R_s (dM/M) \leq dR_s < R_s (dM/M) \text{, or}$$

$$dR_s = \gamma R_s (dM/M) \qquad (31),$$

with $\gamma$, as we shall see, in the range $-1.458 < \gamma \leq +1$.

We now have the means to solve for $v(R_s)$ & $\alpha$ introduced in §3. Resuming the problem begun there, in which a polytrope attracts a spherical shell of mass $(1+\alpha)dM$, but in accreting $dM$ sheds its excess-to-the-needs-of-virial-stability contraction energy and scalar momentum by sending $\alpha\,dM$ of the infall back to infinity. The energy equation for this process, after substituting equation 31 into 26, balances the excess contraction energy from accreting $dM$ against the extra energy carried away to infinity by $\alpha \, dM$,

$$\therefore \quad \tfrac{1}{2} dM \lambda_n^2 V_{esc,n}^2 (2 - \gamma) = \tfrac{1}{2} \alpha \, dM \left[ v^2(R_s) - V_{esc,n}^2 \right],$$

where $v(R_s)$ is the velocity of $\alpha \, dM$ as it leaves the polytrope. This simplifies to

$$(2 - \gamma) \lambda_n^2 = \alpha \left[ \frac{v(R_s)}{V_{esc,n}} + 1 \right] \left[ \frac{v(R_s)}{V_{esc,n}} - 1 \right] \quad (32).$$

The momentum equation, on substituting equation 31 into equation 25, is similarly

$$\tfrac{1}{2} \lambda \lambda_n (1 - \gamma) dM V_{esc,n} = \alpha \, dM \left[ v(R_s) - V_{esc,n} \right],$$

which simplifies to

$$\tfrac{1}{2} \lambda \lambda_n (1 - \gamma) = \alpha \left[ \frac{v(R_s)}{V_{esc,n}} - 1 \right] \quad (33).$$

The simultaneous solution of equations 32 & 33 gives

$$v(R_s) = \left[ \frac{2(2 - \gamma) \lambda_n}{\lambda (1 - \gamma)} - 1 \right] V_{esc,n} \quad (34),$$

$$\alpha = \frac{\lambda^2 \lambda_n (1 - \gamma)^2}{4(2 - \gamma) \lambda_n - 4(1 - \gamma) \lambda} \quad (35).$$

In particular the isentropic case ($\gamma = -1/3$) with the isothermal polytrope ($n = 3/2$) has $\alpha = 0.824$ and $v(R_s) = 1.488 \, V_{esc,3/2}$. So this solution requires $\alpha / (1 + \alpha)$ or 45% of the infalling mass to be returned to infinity.

Figure 1 shows the dependence of $\alpha$ and $\xi = v(R_s)/V_{esc}$ on the choice of accretion law signified by $\gamma$. In the limit of $\gamma \to 1$, set by the requirement that $dP$ from equation 25 be positive, so that an accreting polytrope can reach equilibrium by having an exportable excess of scalar momentum, $\alpha \to 0$ and $\xi \to \infty$. At the opposite extreme as $\gamma \to -1.458$, $v(R_s) \to V_{esc}$ and $\alpha \to \infty$. Thus an accreting polytrope needs a very efficient accelerating mechanism acting on the discarded mass to approach the first limit, while it can only approach the second limit by retaining almost none of the infalling mass. Any system that accretes from infinity must therefore have an intermediate value of $\gamma$.

This treatment is restricted to collisionless systems, as these have no means for exporting energy or cancelling momentum, and so *must respond dynamically.* We are thus led to the conclusion that the symmetric accretion of mass is accompanied by an efflux of part of it: systems shed some mass in gaining mass. However the operation of this loss-mode in collisionless systems automatically assures its operation to some degree in normal dissipative systems. It is therefore worth checking on a few of the values of $\gamma$ exhibited by such systems. Many of these lose mass in such a way that their velocity at infinity $V_\infty \approx V_{esc}$ (Livio 2000). Besides being evidence for a rationale of the type adduced here, this datum, on the assumption that the systems accrete from infinity rather than from an accretion disk, imply the operation of accretion laws with $\gamma \sim -0.43$. On the other hand, when $\gamma$ is estimated from the change to $R_s$ induced by adding mass to a white dwarf, we find that (i) $\gamma \to -1/3$ as $M \to 0$; (ii) $\gamma$ increases smoothly to -0.628 as $M \to 0.6\,M_\odot$; (iii) $\gamma$ reaches -1.458 as $M \to 1.027\,M_\odot$; and (iv) $\gamma < -12$ by $M = 1.4\,M_\odot$. These values are calculated from the relation between $M$ and $R_s$ given by Hansen & Kwaler (1994, p127), which derives from fitting the results of numerical models for white dwarfs with an electron mean molecular weight $\mu_e \equiv 2$. By contrast, at the opposite extreme, the mass influx through the Schwarzschild radius of a black hole has $\gamma \equiv 1$, the unique $\gamma$ that allows mass-accretion without producing either an excess or a deficit of scalar momentum. The diversity in the values of $\gamma$ adduced here clearly shows that while its value is set by the particular characteristics of each system, equations 34 & 35 correctly delineate its operative range while systems are remote from any relativistic limit.

## 6. Conclusions

(i) The scalar sum over the momenta of particles in the rest frame of a closed system, while omni-directional in Cartesian space, is a radially-directed quantity in centrally-symmetric space. Moreover *it is a state variable of the system that pairs with its kinetic energy, as it assumes the same value whenever the kinetic energy has the one unique matching value.* It can thus be considered to be the "internal momentum" of the system.

(ii) A polytrope is unstable against spherically-symmetric mass-loss when conservation of energy is erroneously considered to be the only constraint; a polytrope is inherently stable under the joint constraints of conservation of energy and "internal momentum".

(iii) Polytropes, like black holes, have a predictable entropy increase as they increase in mass. However, if the infalling shell is completely absorbed, its entropy content is lost. Consequemtly the only certain way for a system to guarantee that the entropy of the Universe as a whole increases under a point symmetric accretion of mass is for a fraction of the infalling mass to return to infinity. This mass efflux allows the Universe to conserve mass, energy, and "internal momentum". A system must lose some mass to gain mass.

(iv) When the spherically symmetric accretion of mass occurs in the centrally symmetric potential of a collisionless polytrope, the only possible response of the polytrope lies in a radial redistribution of its mass. Moreover, when a relation is specified between the mass and surface radius, both the fraction of the infall-mass returned to infinity and its velocity are set by the dual constraints of conserving energy and "internal momentum".

This work is supported by the National Astronomy and Ionosphere Center, which is operated by Cornell University under a management agreement with the National Science Foundation.

# 7: References

Chandrasekhar, S. 1939, *Stellar Structure* (New York: Dover)

Hansen, C. J., & Kwaler, S. D. 1994, *Stellar Interiors,* (New York: Springer)

Jaynes, E. T. 1957, Phys.Rev., 106, 620

Jaynes, E. T. 1980, in Papers on Probability, Statistics and Statistical Physics ed. R. D. Rosenkrantz (Dordrecht: Reidel) p407

Knapp, G. R., Phillips, T. G., Leighton, R. B., Lo, K. Y., Wannier, P. G., Wootten, H. A., & Huggins, P. J., 1982, ApJ, 252, 616

Landau, L. D., & Lifshitz, E. M. 1960, *Mechanics,* (Pergamon: Oxford)

Lewis, B. M. 1997, ApJ, 491, 846

Livio, M., 2000, in *Cosmic Explosions,* eds. S. S. Holt & W. W. Zhang (Melville: AIP Conf.Proc. 522) p275

Lynden-Bell, D. 1967, MNRAS, 136, 101

Nakamura, T. K. 2000, ApJ 531, 739

Salpeter, E. E. 1974, ApJ, 193, 585

Figure Legend

Figure 1: The variation with $\gamma$ of $\xi = v(R_s)/V_{esc}$ and $\alpha$ (the fraction of accreted mass ejected to infinity being $\alpha/(1+\alpha)$ ) specified by equations 34 & 35. The parameter $\gamma$ links the change in the polytrope's surface radius with the accreted mass, following equation 31.

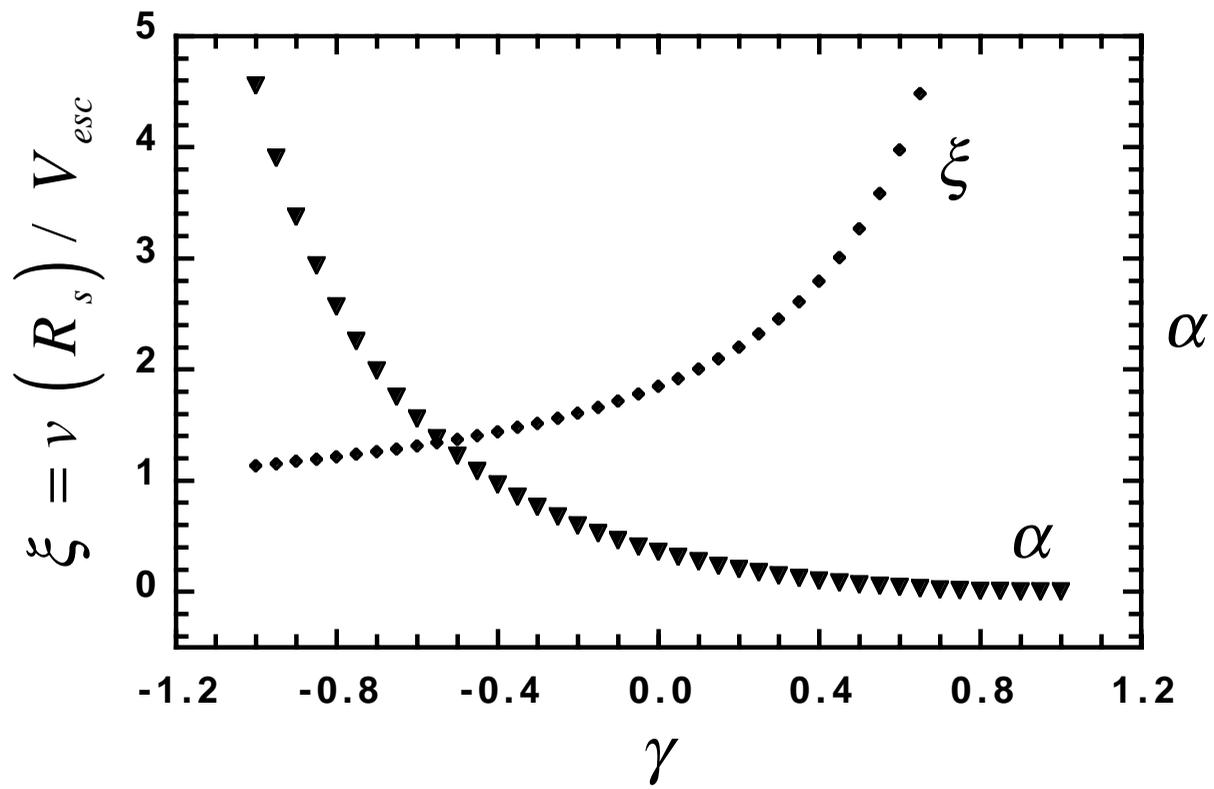